\documentclass{ws-procs961x669}            
\usepackage{hyperref}

\begin{document}
\title{Principles for a Distinguished Global Vacuum:\\
Entropy and the Vacuum State in Causal Set Theory}

\author{Joshua Y. L. Jones}

\address{School of Theoretical Physics, Dublin Institute for Advanced Studies,\\
10 Burlington Road, Dublin 4, Ireland\\
\phantom{line}\\
E-mail: jjones@stp.dias.ie}

\begin{abstract}
Using the framework of real scalar field theory on causal sets, the intimate relation of the Sorkin-Johnston vacuum to entropic purity is elucidated. It is shown that taking a set of sensible principles, and the most natural assumption on the space of fields, leaves the Sorkin-Johnston state as the only candidate for the global vacuum of a quasifree theory.
\end{abstract}


\bodymatter

\section{Introduction}
A theory of quantum gravity should include quantized spacetime. We will take the effective physical manifestation of such a spacetime to be discrete, with a notion of causal order. This causal ordering is sufficient to retrieve the metric of a spacetime up to a conformal factor,\cite{HKM76, Malament77} which discreteness provides.\cite{Riemann1873} This structure is then a causal set,\cite{BLMS87}${}^{,}$\footnote{Due to the brevity of these proceedings, we cannot delve into details regarding the causal set itself. The reader is directed to ref\cite{ButterfieldDowker2024} for an exposition on its genericity, and ref\cite{SuryaReview} for general background.} and we will study its semiclassical interplay with quantum fields, analogous to quantum field theory on curved continuum spacetime. 

We will cover the basic desiderata for an entirely covariant approach to quantum field theory, and how they should be expressed in the causal set. In particular, we will focus on the condition of entropic purity, which must be understood spectrally, and in relation to a spacetime region.\cite{Sorkin2014} We will then see that by taking a set of well motivated principles and a natural field space, we pick out a distinguished global vacuum state, the Sorkin-Johnston state.\cite{Johnston2009, Sorkin2011, AAS2012}

\section{Covariant Quantum Field Theory Preliminaries}

We first mention continuum concepts that will be translated, and used, in the causal set. We take the case of a globally hyperbolic \hyperlink{GH}{(\textbf{GH})} spacetime, and a quasifree \hyperlink{QF}{(\textbf{QF})} real scalar quantum field, $\phi$. Inherent to taking a quasifree scalar field theory are canonical commutation relations \hypertarget{CCR}{(\textbf{CCR})}, and an equation of motion of normally hyperbolic type \hypertarget{NH}{(\textbf{NH})}, which we will denote generally via $\hat{K}\phi = 0$.

\subsection{Wightman Distributions}
Following Wightman, we will treat quantum field theories via their n-point vacuum expectation values, $W(x_{1},\ldots,x_{n}) := \langle 0| \phi(x_{1}),\ldots,\phi(x_{n}) |0 \rangle$. These are generically tempered distributions, but on the causal set, they are only matrices of the distribution kernel restricted to discrete spacetime points. In Minkowski spacetime, they must satisfy a set of axioms,\cite{SW64}
\begin{enumerate}
\renewcommand{\labelenumi}{\theenumi}
\renewcommand{\theenumi}{\textbf{W\arabic{enumi}}}
    \item Positive Semidefiniteness (State norms are non-negative) \label{positivity}
    \item Hermiticity (Inner product conjugate symmetric) \label{hermiticity}
    \item $W(\ldots,x,y,\ldots)=W(\ldots,y,x,\ldots) \quad \forall \text{ spacelike }x,y$ \hyperlink{CCR}{(\textbf{CCR})}\label{spacelike}
    \item Invariant under $\mathcal{P}^{\uparrow}_{+}$ action ($\mathcal{P}^{\uparrow}_{+}$ action unitary)
    \label{LI}
    \item Spectrum condition (Positive energy spectrum):\\
    $\mathcal{F}(W) \neq 0 \, \iff \, \left(\sum_{n} p_{i} = 0 \text{ and } \{p_{i},p_{i}+p_{i+1},\cdots\} \subset V^{+} \right)$, \label{spectrum}
    \item Cluster Property (Vacuum vector is unique): $\forall \text{ spatial vectors } a$ \\
    $\lim_{|a|\to \infty} W(x_{1}+a,\ldots,x_{n}+a,y_{1},\ldots,y_{m})=W(x_{1},\ldots,x_{n})W(y_{1},\ldots,y_{m}). \label{cluster}$
\end{enumerate}
Here $\mathcal{P}^{\uparrow}_{+}$ denotes the proper orthochronous Poincaré group, $\mathcal{F(\cdot)}$ gives the Fourier transform with the $p_{i}$s as momentum space coordinates, and $V^{+}$ is the forward light cone. Given all the Wightman distributions of a theory, one can reconstruct the Hilbert space formulation,\cite{SW64,GN,S} and we have indicated in brackets why each axiom is required. Our quantum field theory is chosen to be quasifree \hypertarget{QF}{(\textbf{QF})}, meaning that n-point distributions are obtained in the same Wickian way as in a free theory. This implies the reconstruction of the theory given only the two point distribution.

The Wightman axioms must be modified to apply to generic curved spacetimes, and there have been many generalisations invoking the Hadamard condition.\cite{Wald77,Wald78,KW91,Radzikowski96,BFK96,BF2000,FV2013,HW2009} Our approach is different to this common one, but not necessarily at odds with it.

\subsection{The Peierls Prescription}

Global hyperbolicity \hypertarget{GH}{(\textbf{GH})} is a constraint upon our spacetime, precluding closed causal curves, and constraining the past and future of any points to have compact intersection. This ensures the existence and uniqueness of retarded and advanced Green distributions, $G^{R}$ and $G^{A}$, for our normally hyperbolic \hyperlink{NH}{(\textbf{NH})} operator. From these, we define the causal propagator, $G := G^{R}-G^{A}$.\footnote{The mathematically inclined reader can consult e.g. ref\cite{BGP} for details about these objects.} Once again, in causal set theory, these are matrices, which are explicitly known in certain cases.\cite{Johnston2008, X2017}

The causal propagator is directly related to the spacetime commutator of the field. This was made precise by Peierls,\cite{Peierls52} and for a quasifree \hyperlink{QF}{(\textbf{QF})} scalar \hyperlink{CCR}{(\textbf{CCR}}, \hyperlink{NH}{\textbf{NH})} theory, we have
\begin{equation}
    i\Delta(x,y) := [\phi(x),\phi(y)] = i G(x,y).
\end{equation}
The object $\Delta$ is called the Pauli-Jordan distribution, and the imposition of its equality with the causal propagator, i.e. $\Delta = G$, will be called \hypertarget{P}{(\textbf{P})}.

\section{Covariant Global Entropy}\label{entropy}
We can now review a formula for a spectral means of calculating entropy for a spacetime, which has the benefit of manifest general covariance, especially in the context of a spacetime regularization.\cite{Sorkin2014} We will take the case of a causal set, but the formula applies for all spacetime regions with an associated countable set of field operators. The value calculated for the entropy is interpretable as that of a hypersurface for which the spacetime is the domain of dependence, if such a hypersurface exists, due to quantum dynamics preserving the algebra.\footnote{This concept often goes by the name of primitive causality. We need not take it as an axiom as it follows from (\textbf{NH}) and (\textbf{GH}).} We start with the spacetime Wightman matrix, with $i$ and $j$ an indexing of the field operators, and decompose it into the anti-commutator and commutator matrices, 
\begin{equation}
    W^{i j} = \langle 0|\phi^{i}\phi^{j}| 0 \rangle = \frac{1}{2}\Bigl(\langle 0 |\{\phi^{i},\phi^{j}\}| 0 \rangle + \langle 0 |[\phi^{i},\phi^{j}]| 0 \rangle\Bigr) =: \frac{1}{2} (H^{i j} + i\Delta^{i j}).
\end{equation}
Here the imposition of hermiticity (\ref{hermiticity}) ensures that $H$ and $\Delta$ are real matrices.

Before we proceed, we must impose that the Hilbert space is formed of only solutions to the equation of motion. To do so, we use the fact that $\text{image}(\Delta)=\text{kernel}(\hat{K})$,\cite{Sorkin2017} and strip out the kernel of $\Delta$ from the space.\footnote{There are indications that one may need to remove even the approximate kernel of $\Delta$, see ref\cite{SY2018}.} We note that the positive semidefiniteness (\ref{positivity}) of $W$, along with the observation that $H$ is symmetric and $\Delta$ is skew-symmetric, implies $\text{kernel}(H) \subseteq \text{kernel}(\Delta)$. We finally remark that this discussion of the kernel of $\hat{K}$ required the equation of motion to hold \hyperlink{NH}{(\textbf{NH})}, and henceforth, we take both the kernels of $H$ and $\Delta$ to have been removed.

With the kernel removed, we can go to a $p$ and $q$ quadrature basis for the modes of $i\Delta$, wherein each mode has a single $p$ and $q$, and there is only one nontrivial commutation relation, $[q^{i},p^{j}] = i \delta^{i j}$. Choosing our basis vector to be $(q_{1},\ldots,q_{n},p_{1},\ldots,p_{n})$, we can use Williamson's theorem to simplify the form of $H$.\footnote{$H$ is now positive definite from the removal of the kernel.} We make a symplectic transformation, which retains the form of $\Delta$, to go to a basis where $H$ is diagonal, with elements given by the positive eigenvalues of $i \Delta^{-1} H$ with doubled multiplicity. Our system now splits up into a series of single mode oscillators, each with a Wightman function given by
\begin{equation} \label{decoupled}
    W_{m} = \frac{1}{2}\left(H_{m}+i\Delta_{m}\right) = \frac{1}{2}\begin{pmatrix}
        \lambda & i \\
        -i & \lambda
    \end{pmatrix}.
\end{equation}
We first note the 2 eigenvalues of $-i\Delta^{-1}_{m} W_{m}$ for each mode, $\frac{1}{2}(1 + \lambda)$ and $\frac{1}{2}(1 - \lambda)$, and label them $\omega_{+}$ and $\omega_{-}$.\footnote{Heretofore, I have not distinguished bilinear forms from operators, meaning the space of fields is $L^{2}$. Although $L^{2}$ is the natural choice, one could add an axiom of field space.\cite{BF13, Sorkin2017}} We can now calculate the entropy. The quasifree nature \hyperlink{QF}{(\textbf{QF})} of our theory constrains the general density matrix for a mode, and as we have the expectation values of all quadratic products of $p$ and $q$ from $W_{m}$, we can explicitly write down this density matrix in the $q$ basis, which has a known entropy.\cite{SorkinEE, SredEE} These are
\begin{gather}
    \rho(q,q') = \sqrt{\frac{\lambda^{-1}}{\pi}}\exp\left(-\frac{\lambda^{-1}}{2}(q^{2}+q'^{2})-\frac{\lambda-\lambda^{-1}}{4}(q-q')^{2})\right), \label{dens1} \\
    S=\frac{\lambda+1}{2}\ln{\frac{\lambda+1}{2}} - \frac{\lambda-1}{2}\ln{\frac{\lambda-1}{2}}.
\end{gather}
This entropy can be written in terms of the eigenvalues of $-i\Delta^{-1}_{m}W_{m}$ as
\begin{equation}
    S=\omega_{+}\ln{(\omega_{+})} + \omega_{-}\ln{(-\omega_{-})}.
\end{equation}
The entropy of the full system is the sum over the decoupled entropies, so we have
\begin{equation}
     S(W) = \sum_{\zeta \in \text{spec}\{Z\}} \zeta \ln{|\zeta|}, \qquad Z := -i\Delta^{-1} W,\label{Seqn}
\end{equation}
using $\lambda \geq 1$ to introduce the absolute magnitude sign. This formula has been used to calculate entanglement entropies in simple causal set spacetimes, with agreement to established continuum hypersurface calculations.\cite{SY2018, KMY2022, DJY2022}

\section{Covariant Distinguished Global Vacuum Construction}

We can now detail the Sorkin-Johnston prescription.\cite{Johnston2009,Sorkin2011,AAS2012} We begin by imposing positive semidefiniteness (\ref{positivity}) and hermiticity (\ref{hermiticity}), the two Wightman axioms that apply straightforwardly to general spacetimes. As we will construct the vacuum for an entire spacetime region, we will impose that the vacuum state is pure. We will also impose that the vacuum state should inherit all the symmetries of the underlying manifold, a generalisation of Poincaré invariance (\ref{LI}). We denote the action of a symmetry transformation of the manifold with $g$.
\begin{enumerate}
\renewcommand{\labelenumi}{\theenumi}
\renewcommand{\theenumi}{\textbf{A\arabic{enumi}}}
    \item ($\equiv$ \ref{positivity}) Positive Semidefiniteness: $W \geq 0$ \label{sjpos}
    \item ($\equiv$ \ref{hermiticity}) Hermiticity: $W = W^{\dag}$ \label{sjherm}
    \item \phantom{($\equiv$ \ref{hermiticity})} Purity: $S(W)=0$ \label{sjpure}
    \item {($\cong$ \ref{LI})} Manifold Symmetry: $g \circ W= W$ \label{sjsym}
\end{enumerate}
For now, we take the first three axioms and proceed; we will return to discuss the action of ($\ref{sjsym}$) later. The question of the field space remains open, so we furnish our space with the $L^{2}$ inner product. Were a different inner product taken, a similar procedure would apply. The primary relevance of the inner product is in the entropy equation, equation (\ref{Seqn}), that we now use to impose purity (\ref{sjpure}). For $S$ to vanish, the eigenvalues of $Z := -i\Delta^{-1} W$ must be $0$ or $1$. This means $Z$ is a projection operator, i.e. $Z^{2}=Z$, and
\begin{equation}
    W\Delta^{-1}W = iW.
\end{equation}
Using (\ref{sjherm}) and the decoupled basis of (\ref{decoupled}), we see $H$ and $\Delta$ commute, and are simultaneously diagonalisable. This implies we have $H^{2}=-\Delta^{2}$, and thus
\begin{equation}
    H = \pm \sqrt{-\Delta^{2}},\label{befbranch}
\end{equation}
where the square root acts elementwise on the spectrum of $-\Delta^{2}$.\footnote{It is known that $\Delta$ is bounded and symmetric for the $L^{2}$ inner product on a bounded spacetime.\cite{FV2012} For unbounded spacetimes, we understand the above procedure as the result of a limiting process.} Imposing the axiom of positive semidefiniteness (\ref{sjpos}), we thus have
\begin{equation}
    W = \frac{1}{2}\left( \sqrt{-\Delta^{2}} + i\Delta \right) = \text{pos}(i\Delta).
\end{equation}
We can now see how ($\ref{sjsym}$) came into our vacuum, which is via the inner product. Were our inner product not invariant under the symmetry action, the resultant vacuum state would similarly not be invariant, and so the imposition of ($\ref{sjsym}$) is realised as the choice of an inner product that is invariant under the manifold symmetries. Unfortunately, this condition does not constrain the inner product entirely, so in the absence of more guiding principles, we stick to our choice of $L^{2}$, which is guaranteed to be invariant under symmetry transformations for all spacetime manifolds. 

We see that after imposing global hyperbolicity \hyperlink{GH}{(\textbf{GH})}, normal hyperbolicity \hyperlink{NH}{(\textbf{NH})}, and quasifreedom \hyperlink{QF}{(\textbf{QF})}, $W$ specifies the entire quantum field theory, after we supply $\Delta$ via the Peierls relation \hyperlink{P}{(\textbf{P})}. We can thus regard the Sorkin-Johnston prescription as the result of imposing seven conditions upon the vacuum state, and the choice of $L^{2}$ field space (this choice being constrained by the condition ($\ref{sjsym}$)),
\begin{equation}
    \ref{sjpos} \land \ref{sjherm} \land \ref{sjpure} \land \hyperlink{GH}{(\textbf{GH})} \land \hyperlink{NH}{(\textbf{NH})} \land \hyperlink{QF}{(\textbf{QF})} \land \hyperlink{P}{(\textbf{P})} \land \mathbf{L^{2}}(\ref{sjsym}) \, \to \, W = \text{pos}(i (G^{R}-G^{A})).
\end{equation}
Let us take a moment to note that \hyperlink{CCR}{(\textbf{CCR})} was used to get \hyperlink{P}{(\textbf{P})}. Our construction was under the supposition of a scalar field theory, but a modified \hyperlink{P}{(\textbf{P})}, and full set of principles, should apply just as well for fermions.

Having constructed our field theory, some comments are in order pertaining to the non-locality of the construction. There is a no-go theorem, by Fewster and Verch, precluding a distinguished vacuum choice for what they call a ``dynamically local" theory.\cite{FVnogo} It is clear that our construction evades this theorem via being non-local, as the $L^{2}$ inner product on the entire spacetime is used to determine purity. In general, the restriction of the vacuum state of a quantum field to a subregion of a spacetime should not be pure, whilst purity is a fundamental tenet of the Sorkin-Johnston perscription. One should understand the Sorkin-Johnston prescription as applicable only to complete spacetimes, for which vacuum states should be pure.

Whilst we may accept some degree of non-locality, we should still consider the possibility of faster than light communication. It is clear that we have precluded the possibility of signalling superluminally directly via the field, via \hyperlink{P}{(\textbf{P})}, but it may seem that one can signal arbitrarily fast via manipulating the spacetime gravitationally, to influence the vacuum state. We have spooky action at a distance.

Of course, this scenario is impossible in the context of quantum field theory in curved spacetime, due to the fixed nature of the background spacetime. This pathology is the same as that present in all global vacuum assignments within the framework. The axiom of manifold symmetry (\ref{sjsym}), included in conventional vacuum assignments, is fundamentally non-local. All it suggests is that when one goes to a dynamical spacetime, one should find a causal way to evolve the vacuum, and that this non-locality is a feature of vacuum assignment on a fixed background.

\subsection{Relation to the Wightman Axioms}
In our construction of the state, we have explicitly used three generalised Wightman axioms (\ref{sjpos},\ref{sjherm},\ref{sjsym}). Let us now consider the others.

The axiom for symmetry under spacelike interchange (\ref{spacelike}) is satisfied noting that (\ref{sjherm}) and \hyperlink{P}{(\textbf{P})} impose that the antisymmetric part of $W$ is $i \Delta$, which vanishes for spacelike points due to the causal nature of the Green functions. The cluster property (\ref{cluster}) is satisfied by $W(x,y)$ vanishing for infinitely spacelike separated points. This is easiest to see by reinserting coordinate labels into equation (\ref{befbranch}), such that we have
$H(x,z) = \pm \sqrt{\sum_{y} \Delta(x,y) \Delta(y,z)},$ which acts as an operator via summation over the $z$ coordinate. As we take $x$ and $z$ to be infinitely spacelike separated, the causal support of the Green kernels means they must have disjoint supports, so $\lim_{a\to\infty}H(x,x+a)=0$, which implies $\lim_{a\to\infty}W(x,x+a)=0$, recovering the cluster property. This is a consequence of (\ref{sjherm}) and (\ref{sjpure}), coupled with the support of $\Delta$ imposed by \hyperlink{P}{(\textbf{P})}.

The final axiom is the spectrum condition (\ref{spectrum}). For static spacetimes, i.e. ones with a hypersurface-orthogonal timelike Killing field, we may use this Killing field to decompose frequency, and impose that the vacuum state contains only positive frequencies. In general spacetimes, no such Killing field exists, and we require a different technique. The Sorkin-Johnston prescription suggests that we replace the notion of positive frequency with the positive eigenspectrum of the commutator.

This conclusion was the result of theorising our way to a vacuum state via the imposition of desirable properties, but to be a generalisation for vacuum state selection, the Sorkin-Johnston state must reproduce uncontroversial vacuum states, such as those of the aforementioned static spacetimes. In the case of static spacetimes of infinite timelike extent, it is known that the Sorkin-Johnston state always agrees with the conventional notion of frequency splitting.\cite{AAS2012}${}^{,}$\footnote{For ultrastatic slab spacetimes of finite timelike extent, the Sorkin-Johnston state is not generically Hadamard,\cite{FV2012} though the Hadamard property may be obtained by changing the field space.\cite{BF13} Further study should reveal whether field space modification is necessary in physical spacetimes, and how the Sorkin-Johnston state's Hadamardness is affected by geodesic incompleteness.\cite{FV2013}} There have also been studies of the Sorkin-Johnston state in other spacetimes, most interestingly in the case of de Sitter spacetime. There its exact properties are unclear, and there are indications that it may not be Hadamard, or somehow not even one of the $\alpha$-vacua.\cite{AB13, SXY19} This is an exciting direction for further investigation, as such simple non-static spacetimes should be key to understanding the field space.

\section*{Acknowledgements}
I would like to thank Yasaman Yazdi for many discussions, and suggestions on the direction of these proceedings. I am supported by Science Foundation Ireland under Grant number 22/PATH-S/10704.

\bibliographystyle{ws-procs961x669}
\bibliography{ws-pro-sample}

\end{document}